\documentclass[aps,pra,epsfig,graphics,floatfix,mathbbm,twocolumn,,a4paper]{revtex4}
\usepackage{amsmath,amsfonts,amssymb,graphics,graphicx,epsfig,color,times,bbm}

\begin{document}
\bibliographystyle{apsrev}
\title{Entangling capacity and distinguishability of two-qubit unitary operators}
\author{Anthony Chefles}
\affiliation{Department of Mathematical Physics, NUI Maynooth,
Maynooth, Co. Kildare, Ireland.} \email{achefles@thphys.nuim.ie}

\begin{abstract}
\vspace{0.5cm} We prove that the entangling capacity of a two-qubit
unitary operator without local ancillas, both with and without the
restriction to initial product states, as quantified by the maximum
attainable concurrence, is directly related to the
distinguishability of a closely related pair of two-qubit unitary
operators. These operators are the original operator transformed
into its canonical form and the adjoint of this canonical form. The
distinguishability of these operators is quantified by the minimum
overlap of the output states over all possible input probe states.
The entangling capacity of the original unitary operator is
therefore directly related to the degree of non-Hermiticity of its
canonical form, as quantified in an operationally satisfactory
manner in terms of the extent to which it can be distinguished, by
measurement, from its adjoint. Furthermore, the maximum entropy of
entanglement, again without local ancillas, that a given two-qubit
unitary operator can generate, is found to be closely related to the
classical capacities of certain quantum channels.

 \vspace{0.5cm} .

\end{abstract}
\pacs{03.67.Hk, 03.67.Mn} \maketitle
\section{Introduction}
\label{sec:1}
\renewcommand{\theequation}{1.\arabic{equation}}
\setcounter{equation}{0}

An interesting development that has recently taken place in quantum
information science has been the widespread realisation that the
information-theoretic properties of quantum operations are of
comparable importance to those of quantum states. We may view this
to be a consequence of the fact that all properties of quantum
states, including those most relevant to quantum communications and
computation, such as entanglement and nonorthogonality, must somehow
be created.  It is quantum operations that are responsible for the
creation of states with these properties. As such, there has been a
considerable amount of activity devoted to establishing
relationships between properties of quantum states, especially those
that are useful in applications, and the properties of quantum
operations that give rise to them.

Perhaps the most important resource in quantum communications and
computation is entanglement.  Entanglement is a resource that must
be created by operations.  Consequently, a considerable effort has
been directed at characterising the ability of quantum operations to
generate entanglement
\cite{KC,KG,Makhlin,LHL,ZVWS,Rezakhani,Zanardi1,Zanardi2,Zanardi3}.
Of particular importance is the maximum amount of entanglement that
a quantum operation can generate for some set of initial states.
Such a quantity is known as an entangling capacity.

The general nonorthogonality of quantum states is also one of
their most pertinent and useful features.  Nonorthogonal states
cannot be perfectly distinguished.  This fact forms the basis for
many quantum cryptographic protocols.  The distinguishability of
operations is closely related to the distinguishability of the
states they produce, since it is by discriminating among the
output states of quantum operations that we distinguish among the
operations themselves \cite{CPR,Acin,DPP,retro}.

The aim of this paper is to prove and explore the consequences of a
curiously simple relationship between the entangling capacity of and
the distinguishability of bipartite unitary operations acting on two
qubits.  Specifically, we make use of the fact that any unitary
operator $U_{AB}$ on two qubits $A$ and $B$ can, using local
operations on $A$ and $B$, be put into a certain canonical form
$U_{d}$. We find that the entangling capacity of $U_{AB}$, as
quantified by the maximum concurrence of the output states it
generates, with or without the restriction to an initial product
state, but without ancillas, is directly related to the
distinguishability of $U_{d}$ and $U_{d}^{\dagger}$. Here,
distinguishability is quantified by the minimum, over all initial
states, of the overlap of the output states they create.  This
result extends one previously obtained by Zhang et al. \cite{ZVWS}
which relates to perfect entanglers.

In Section \ref{sec:2}, we review the relevant background material
on the entangling capacity and distinguishability of unitary
operators.  Section \ref{sec:3} is devoted to proving our main
result relating to maximum concurrence and distinguishability and
to exploring some of its consequences.

For bipartite pure states, the entropy of entanglement is often
used as an entanglement measure.  It is interesting to enquire as
to whether or not the entangling capacity, in terms of the maximum
entropy of entanglement, also has a direct operational connection
with the distinguishability of states.

The entropy of entanglement draws its significance from its
asymptotic properties, when there are many copies of a given
entangled state available.  It is natural to suspect that the
entropic entangling capacity of a given quantum operation may
relate to the distinguishability of quantum operations when the
operation is performed many times.  This leads us to consider
classical information theory, which is of an intrinsically
asymptotic nature.

In Section \ref{sec:4}, we confirm this suspicion by showing that
the entropic entangling capacity of $U_{AB}$ is related to
classical information transmission in two distinct ways.
 The first relationship concerns the so-called first-order
classical capacity, where collective decoding of the individual
signal carriers is forbidden.  The second relates to the
Hausladen-Jozsa-Schumacher-Westmoreland-Wootters (HJSWW) capacity
\cite{HJSWW}, where arbitrary collective decoding of the pure state
signals is permitted.  The relationship we describe relates to the
transmission of pure states.  The generalisation of the HJSWW
capacity to mixed state signals is the well-known
Holevo-Schumacher-Westmoreland capacity \cite{Holevo,SW}.  We
therefore find that the relationship between the entangling capacity
and distinguishability is not simply of a ``one-shot" nature, but
also has some interesting implications for the asymptotic limit. We
conclude in Section \ref{sec:5} with a general discussion of our
results.

\section{Entangling capacities and distinguishability of unitary operators}
\label{sec:2}
\renewcommand{\theequation}{2.\arabic{equation}}
\setcounter{equation}{0}
\subsection{Entangling capacities}
 Consider two quantum systems $A$ and $B$. Associated with each system is a copy of the Hilbert space ${\cal H}$.  The
 composite system has Hilbert space ${\cal H}_{AB}={\cal H}^{{\otimes}2}$.  We shall take
$A$ and $B$ to be qubits and so ${\cal H}$ is two-dimensional.

Let us consider a unitary operator $U_{AB}$  on ${\cal H}_{AB}$.
We would like to know how much entanglement such an operator can
create between these two systems.  To address this issue, we must
choose a measure of entanglement.  For a  bipartite pure state of
two qubits, two particular entanglement measures are in common
use. These are the concurrence and the entropy of entanglement.

Let $|{\Psi}{\rangle}{\in}{\cal H}_{AB}$ be an arbitrary pure
state of $AB$.  Let ${\rho}_{A}^{\Psi}$ (${\rho}_{B}^{\Psi}$) be
the corresponding reduced density operator of $A$ ($B$) for this
state.  Both of these operators have the same eigenvalues, which
we shall denote by $q_{j}$, where $j=1,2$.  The concurrence of
$|{\Psi}{\rangle}$ may be written as
\begin{equation}
C(|{\Psi}{\rangle})=2\sqrt{q_{1}q_{2}}=2\sqrt{\det({\rho}^{\Psi}_{j})}.
\end{equation}
The entropy of entanglement of $|{\Psi}{\rangle}$ is
\begin{equation}
E(|{\Psi}{\rangle})=-\sum_{j=1}^{2}q_{j}{\log}q_{j},
\end{equation}
where here, as throughout this paper, the logarithm has base 2.
Hill and Wootters \cite{HW} noted that the entropy of entanglement
may be written in terms of the concurrence as
\begin{equation}
\label{entropcab}
E(|{\Psi}{\rangle})=h\left(\frac{1+\sqrt{1-[C(|{\Psi}{\rangle})]^{2}}}{2}\right),
\end{equation}
where $h(x)$ is the binary entropy function:
\begin{equation}
h(x)=-x{\log}x-(1-x){\log}(1-x).
\end{equation}
 Both of these entanglement measures attain their
maximum value, which in both cases is 1, iff $|{\Psi}{\rangle}$ is
maximally entangled. They also take their minimum value, of 0, iff
$|{\Psi}{\rangle}$ is a product state. These entanglement measures
are also invariant under local unitary operations. Finally, they
are monotonically increasing functions of each other, so the
maximization of one is equivalent to the maximization of the
other.  As a consequence of this equivalence we shall, for the
time being, concentrate on the concurrence as an entanglement
measure.

The entangling capacity of a unitary operator is the maximum amount
of entanglement that the operator can generate.  In the most general
situation we can contemplate, $A$ and $B$ can be entangled with
local ancillary systems.  At the time of writing, the most general
solution to this problem is not known.  We may consider instead the
situation where the initial state is a pure state
$|{\Psi}{\rangle}{\in}{\cal H}_{AB}$.  Leifer et al. \cite{LHL}
solved this problem and obtained the general value of
\begin{equation}
\label{entcap1} C_{max}(U_{AB})=\max_{|{\Psi}{\rangle}{\in}{\cal
H}_{AB}}[C(U_{AB}|{\Psi}{\rangle})-C(|{\Psi}{\rangle})].
\end{equation}
In a prior work, Kraus and Cirac \cite{KC} addressed the more
restricted problem where the set of possible initial states was
taken to be the set of two-qubit product states in ${\cal H}_{AB}$.
In this case, writing a typical product state
$|{\Psi}{\rangle}{\in}{\cal H}_{AB}$ as
$|{\Psi}{\rangle}=|{\psi}_{A}{\rangle}{\otimes}|{\psi}_{B}{\rangle}$,
the resulting product entangling capacity may be written as
\begin{equation}
\label{entcap2}
C_{max}^{prod}(U_{AB})=\max_{|{\psi}_{A}{\rangle},|{\psi}_{B}{\rangle}{\in}{\cal
H}}C(U_{AB}|{\psi}_{A}{\rangle}{\otimes}|{\psi}_{B}{\rangle}).
\end{equation}
Kraus and Cirac obtained an explicit expression for
$C_{max}^{prod}(U_{AB})$ for an arbitrary two-qubit unitary
operator. The results of the above authors can be summarized as
follows. Firstly it is known \cite{KC,KG} that any unitary
operator $U_{AB}$ acting on two qubits, $A$ and $B$, can be
written in the form
\begin{equation}
\label{con}
U_{AB}=(X_{A}{\otimes}X_{B})U_{d}(Y_{A}{\otimes}Y_{B}),
\end{equation}
where $X_{A},X_{B},Y_{A}$ and $Y_{B}$ are single-qubit unitary
operators and the bipartite unitary operator $U_{d}$ has the form
\begin{equation}
\label{Ud}
U_{d}={\exp}\left[-i({\alpha}_{x}{\sigma}_{x}{\otimes}{\sigma}_{x}+{\alpha}_{y}{\sigma}_{y}{\otimes}{\sigma}_{y}+{\alpha}_{z}{\sigma}_{z}{\otimes}{\sigma}_{z})\right].
\end{equation}
Here, ${\sigma}_{x},{\sigma}_{y}$ and ${\sigma}_{z}$ are the usual
Pauli spin operators and the vector
$d=({\alpha}_{x},{\alpha}_{y},{\alpha}_{z})$ has real components
satisfying
\begin{equation}
\label{alphaorder0}
0{\leq}|{\alpha}_{z}|{\leq}{\alpha}_{y}{\leq}{\alpha}_{x}{\leq}{\pi}/4.
\end{equation}
For the purposes of this paper, it is sufficient to confine our
attention here to the case of ${\alpha}_{z}{\geq}0$.  We discuss
this matter in more detail in the Section \ref{sec:3} and in the
Appendix. We then have
\begin{equation}
\label{alphaorder1}
0{\leq}{\alpha}_{z}{\leq}{\alpha}_{y}{\leq}{\alpha}_{x}{\leq}{\pi}/4.
\end{equation}
We can write the eigenvalues of $U_{d}$ in the form
$e^{-i{\lambda}_{j}}$, where $j=1,{\ldots},4$.  The
${\lambda}_{j}$ are given by
\begin{eqnarray}
\label{lambdaeq1}
{\lambda}_{4}&=&{\alpha}_{x}+{\alpha}_{y}-{\alpha}_{z}, \\
\label{lambdaeq2}
{\lambda}_{3}&=&{\alpha}_{x}-{\alpha}_{y}+{\alpha}_{z}, \\
\label{lambdaeq3}
{\lambda}_{2}&=&-{\alpha}_{x}+{\alpha}_{y}+{\alpha}_{z}, \\
\label{lambdaeq4}
{\lambda}_{1}&=&-{\alpha}_{x}-{\alpha}_{y}-{\alpha}_{z}.
\end{eqnarray}
It follows readily from these equations and inequality
\eqref{alphaorder1} that the ${\lambda}_{j}$ are ordered according
to
\begin{equation}
\label{lambdaorder}
{\lambda}_{4}{\geq}{\lambda}_{3}{\geq}{\lambda}_{2}{\geq}{\lambda}_{1}.
\end{equation}
In terms of the above definitions, Kraus and Cirac and Leifer et al.
obtained the general forms of $C_{max}^{prod}(U_{AB})$ and
$C_{max}(U_{AB})$ for two-qubit unitary operators. They found that
when the following two inequalities are satisfied,
\begin{eqnarray}
\label{alphaineq1}
{\alpha}_{x}+{\alpha}_{y}&{\geq}&{\pi}/4, \\
\label{alphaineq2} {\alpha}_{y}+{\alpha}_{z}&{\leq}&{\pi}/4,
\end{eqnarray}
then these entangling capacities are
\begin{equation}
\label{kceqn0} C_{max}^{prod}(U_{AB})=C_{max}(U_{AB})=1.
\end{equation}
This implies that $U_{AB}$ can transform some product state into a
maximally entangled state, in which case $U_{AB}$ is said to be a
perfect entangler. When either of these inequalities is violated,
$U_{AB}$ is not a perfect entangler and
\begin{equation}
\label{kceqn1}
C_{max}^{prod}(U_{AB})=[C_{max}(U_{AB})]^{2}=\max_{j,j'}|{\sin}({\lambda}_{j}-{\lambda}_{j'})|.
\end{equation}
Notice that these two inequalities cannot be violated
simultaneously, as this would contradict the ordering of the
${\alpha}_{j}$ in Eq. \eqref{alphaorder1}.

 It is clear from Eqs. \eqref{kceqn0} and \eqref{kceqn1} that the
entangling capacity without the restriction to product states is
always equal to the square root of the entangling capacity with this
restriction.  So, these capacities are very easily deduced from each
other.  For this reason and also because of the fact that, for the
purposes of deriving the principal result of this paper, the product
entangling capacity will be slightly easier to work with, we shall
focus mainly on this quantity in subsequent sections.

We will show that there is an intriguing relationship between the
product entangling capacity \eqref{kceqn1} and distinguishability of
unitary operators. Prior to describing this relationship, we shall
review some facts about the latter topic.

\subsection{Distinguishability}
 Consider two unitary operators $S$ and $T$, referring to a possibly composite system with total Hilbert space ${\cal
H}_{tot}$.  This space is taken to have finite dimensionality $D$.
We aim to distinguish as well as possible between these two
operators using a probe state.  The problem of optimally
discriminating between two unitary operators was first addressed and
solved by Childs et al. \cite{CPR}.   Useful further insights into
this problem were obtained in the subsequent investigations of
Ac\'{\i}n \cite{Acin} and D'Ariano et al. \cite{DPP}.  In
particular, the latter authors established that for optimal
discrimination, the initial probe state may be taken to be a pure
state which is not entangled with any ancillary systems.

\begin{figure}
\begin{center}
\epsfxsize8cm \centerline{\epsfbox{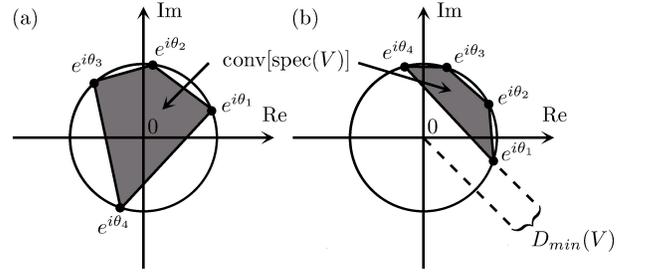}}
\end{center}
\caption{Geometric depiction of (a) perfect and (b) imperfect
distinguishability of two unitary operators, $S$ and $T$. The
$e^{i{\theta}_{j}}$ are the eigenvalues of $V=S^{\dagger}T$. The
set of possible values of the inner product between the final
states $S|{\Phi}{\rangle}$ and $T|{\Phi}{\rangle}$ is, in each
case, $\mathrm{conv}[\mathrm{spec}(V)]$.   The minimum overlap
between these final states is $D_{min}(V)$, the minimum distance
from $0$ to $\mathrm{conv}[\mathrm{spec}(V)]$.  If, as is the case
in (a), $0{\in}\mathrm{conv}[\mathrm{spec}(V)]$, then
$D_{min}(V)=0$ and perfect discrimination between $S$ and $T$ is
possible.  If, on the other hand,
$0{\not\in}\mathrm{conv}[\mathrm{spec}(V)]$, which is the case in
(b), then $S$ and $T$ cannot be perfectly discriminated although
their distinguishability continues to be governed by
$D_{min}(V)$.} \label{figure1}
\end{figure}

In the light of this, let the probe state be some pure state
$|{\Phi}{\rangle}{\in}{\cal H}_{tot}$. To optimally distinguish
between $S$ and $T$, we require that the overlap between the states
$S|{\Phi}{\rangle}$ and $T|{\Phi}{\rangle}$ be as small as possible.
It is convenient to define the unitary operator $V=S^{\dagger}T$ and
to consider the inner product between the final states
\begin{equation}
{\langle}{\Phi}|S^{\dagger}T|{\Phi}{\rangle}={\langle}{\Phi}|V|{\Phi}{\rangle}.
\end{equation}
Since $V$ is unitary, it can be spectrally decomposed as
\begin{equation}
V=\sum_{j=1}^{D}e^{i{\theta}_{j}}|v_{j}{\rangle}{\langle}v_{j}|,
\end{equation}
where the angles ${\theta}_{j}$ are real and the $|v_{j}{\rangle}$
form an orthonormal basis for ${\cal H}_{tot}$. The state
$|{\Phi}{\rangle}$ can be expanded in terms of this basis as
\begin{equation}
|{\Phi}{\rangle}=\sum_{j=1}^{D}c_{j}|v_{j}{\rangle}.
\end{equation}
For the sake of notational convenience, let us define
\begin{equation}
p_{j}=|c_{j}|^{2}.
\end{equation}
It is clear that the $p_{j}$ may take any non-negative values
subject to the normalisation of $|{\Phi}{\rangle}$, which is
equivalent to
\begin{equation}
\sum_{j=1}^{D}p_{j}=1.
\end{equation}
 Combining the above expressions, we obtain
\begin{equation}
{\langle}{\Phi}|V|{\Phi}{\rangle}=\sum_{j=1}^{D}p_{j}e^{i{\theta}_{j}}.
\end{equation}
In discriminating between two pure states, it is their overlap,
rather than their inner product, that is significant.  If we wish to
obtain the minimum value of the overlap of $S|{\Phi}{\rangle}$ and
$T|{\Phi}{\rangle}$, we must evaluate
\begin{equation}
\label{minoverlap}
\min_{|{\Phi}{\rangle}:\||{\Phi}{\rangle}\|=1}|{\langle}{\Phi}|V|{\Phi}{\rangle}|=\min_{p_{j}:\sum_{j=1}^{D}p_{j}=1}\left|\sum_{j=1}^{D}p_{j}e^{i{\theta}_{j}}\right|.
\end{equation}
The above expression has a simple geometrical interpretation that
is depicted  Figure \ref{figure1}. For each particular set of
$p_{j}$, the number $\sum_{j=1}^{D}p_{j}e^{i{\theta}_{j}}$
represents a point in the complex plane.  The set of all such
points is easily recognizable as the convex hull of the
eigenvalues $e^{i{\theta}_{j}}$ \cite{Rockafellar}.  This is the
smallest convex set which contains all of these eigenvalues. Here,
the number of eigenvalues is finite, so their convex hull is
simply a convex polygon with these eigenvalues at its vertices.
For the sake of notational convenience, we shall write the convex
hull of the spectrum of an operator $V$ as
$\mathrm{conv}[\mathrm{spec}(V)]$.

The minimum overlap in Eq. \eqref{minoverlap} is then simply the
minimum modulus of the complex numbers in
$\mathrm{conv}[\mathrm{spec}(V)]$.  In other words, it is the
minimum  distance from 0 to $\mathrm{conv}[\mathrm{spec}(V)]$. Let
us denote this minimum distance by $D_{min}(V)$.  Then we may
write
\begin{equation}
\label{dmin}
\min_{|{\Phi}{\rangle}:\||{\Phi}{\rangle}\|=1}|{\langle}{\Phi}|S^{\dagger}T|{\Phi}{\rangle}|=D_{min}(V).
\end{equation}

For the operators $S$ and $T$ to be perfectly distinguishable for
 some initial probe state $|{\Phi}{\rangle}$, we require that the final
states be orthogonal.  From the above considerations, we see that
this will be the case iff $D_{min}(V)=0$, which is to say that
$0{\in}\mathrm{conv}[\mathrm{spec}(V)]$.

\section{Maximum concurrence and minimum overlap}
\label{sec:3}
\renewcommand{\theequation}{3.\arabic{equation}}
\setcounter{equation}{0}

We shall now show that for any two-qubit unitary operator
$U_{AB}$, the product entangling capacity in Eqs. \eqref{kceqn0}
and \eqref{kceqn1} and the distinguishability of $U_{d}$ and its
adjoint $U_{d}^{\dagger}$, as quantified by the minimum overlap,
satisfy a curiously simple relationship.  The distinguishability
of $U_{d}$ and its adjoint $U_{d}^{\dagger}$ is characterised
using Eq. \eqref{dmin}, where we make the identifications
$S=U_{d}^{\dagger}$, $T=U_{d}$, $V=U_{d}^{2}$ and ${\cal
H}_{tot}={\cal H}_{AB}$.  We are now
in a position to prove our main result.\\

\noindent {\bf{Theorem}}.  For any two-qubit unitary operator
$U_{AB}$ with canonical form $U_{d}$ defined by Eqs. \eqref{Ud}
and \eqref{alphaorder0},
\begin{equation}
\label{maintheorem}
[C_{max}^{prod}(U_{AB})]^{2}+[D_{min}(U_{d}^{2})]^{2}=1.
\end{equation}
To prove this result, we shall treat separately the cases of
$U_{AB}$ being a perfect entangler and not being a perfect
entangler.

Prior to proving Eq. \eqref{maintheorem} for perfect entanglers,
we note that for such operators, this theorem has been effectively
established in Theorem 1 of \cite{ZVWS}, although without
reference to the distinguishability of unitary operators.  Our
discovery of the general validity of Eq. \eqref{maintheorem} came
about through the realisation that the results of these authors
relate to the distinguishability of unitary operators.  This led
us to enquire as to whether or not there is a general relationship
between the distinguishability of $U_{d}$ and $U_{d}^{\dagger}$
and the entangling capacity of ${U}_{AB}$. This enquiry led to our
discovery of the general validity of Eq. \eqref{maintheorem}.
Also, for the case of perfect entanglers, our proof is possibly
slightly simpler.

Another point is that, here, we explicitly prove Eq.
\eqref{maintheorem} only for  cases where ${\alpha}_{z}{\geq}0$.
We do this because the validity of Eq. \eqref{maintheorem} for
negative  ${\alpha}_{z}$ follows readily from its validity for
${\alpha}_{z}{\geq}0$.  We prove this in the Appendix.   \\

\noindent {\bf{Proof}}.  Case (a):  $U_{AB}$ is a perfect entangler.\\

In this case, $C_{max}^{prod}(U_{AB})=1$.  Proving
\eqref{maintheorem} in this case then amounts to showing that
$D_{min}(U_{d}^{2})=0$, implying that $U_{d}$ and
$U_{d}^{\dagger}$ are perfectly distinguishable.

To begin, we know from the preceding section that
$C_{max}^{prod}(U_{AB})=1$ is equivalent to inequalities
\eqref{alphaineq1} and \eqref{alphaineq2} being satisfied. Also,
to have $D_{min}(U_{d}^{2})=0$, we require that
$0{\in}\mathrm{conv}[\mathrm{spec}(U_{d}^{2})]$.  This latter
condition can be understood in simple geometrical terms. Consider
the angular separations of neighbouring eigenvalues of
$U_{d}^{2}$.  To calculate these spacings, it is convenient to
define $\tilde{\lambda}_{j}={\lambda}_{j}-{\lambda}_{1}$, which
all lie in the interval $[0,{\pi}]$. The spacings between
neighbouring angles $\tilde{\lambda}_{j}$ are identical to those
of the original $\tilde{\lambda}_{j}$. In particular, we have
\begin{equation}
\label{tildelambdaorder}
{\pi}{\geq}{\tilde\lambda}_{4}{\geq}{\tilde\lambda}_{3}{\geq}{\tilde\lambda}_{2}{\geq}{\tilde\lambda}_{1}=0.
\end{equation}
Using these transformed angles, it is easy to see that the four
spacings are given by
$2({\tilde\lambda}_{j+1}-{\tilde\lambda}_{j})$ for
$j=1,{\ldots},3$ and $2({\pi}-{\tilde\lambda}_{4})$ for $j=4$.
Using Eqs. \eqref{lambdaeq1}-\eqref{lambdaeq4}, we can write these
spacings as

\begin{eqnarray}
\label{tild1}
2(\tilde{\lambda}_{2}-\tilde{\lambda}_{1})&=&4({\alpha}_{y}+{\alpha}_{z}){\leq}{\pi},
\\
\label{tild2}
2(\tilde{\lambda}_{3}-\tilde{\lambda}_{2})&=&4({\alpha}_{x}-{\alpha}_{y}){\leq}{\pi},
\\
\label{tild3}
2(\tilde{\lambda}_{4}-\tilde{\lambda}_{3})&=&4({\alpha}_{y}-{\alpha}_{z}){\leq}{\pi},
\\
\label{tild4}
2({\pi}-{\tilde\lambda}_{4})&=&2({\pi}-2({\alpha}_{x}+{\alpha}_{y})){\leq}{\pi}.
\end{eqnarray}

The inequalities in \eqref{tild1} and \eqref{tild4} are
consequences of \eqref{alphaineq2} and \eqref{alphaineq1}
respectively, while those in \eqref{tild2} and \eqref{tild3}
follow from \eqref{alphaorder1}.  Now, zero is a element of
$\mathrm{conv}[\mathrm{spec}(U_{d}^{2})]$ iff the eigenvalues of
$U_{d}^{2}$ do not all lie in some arc of the unit circle which
subtends an angle of less than ${\pi}$ radians.  This is
equivalent to the four angular separations being no greater than
${\pi}$ radians, which is what we have just demonstrated.  So, we
have shown that whenever $U_{AB}$ is a perfect entangler, $U_{d}$
and $U_{d}^{\dagger}$
are perfectly distinguishable, and so Eq. \eqref{maintheorem} holds in this case.\\

\noindent Case (b):  $U_{AB}$ is not a perfect entangler.\\

Let us now show that Eq. \eqref{maintheorem} also holds when
$U_{AB}$ is not a perfect entangler.  We will be concerned here
with the minimum distance from the origin to the convex hull of
the complex numbers $e^{-2i{\lambda}_{j}}$.  This distance is
clearly identical to that between the origin and the convex hull
of the $e^{-2i{\tilde\lambda}_{j}}$ and again it will be more
convenient to work with the latter quantities.

In terms of these angles, we may write Eq. \eqref{kceqn1} as
\begin{equation}
\label{kceqn2}
C_{max}^{prod}(U_{AB})=\max_{j,j'}|{\sin}(\tilde{\lambda}_{j}-\tilde{\lambda}_{j'})|.
\end{equation}
We know that $U_{AB}$ is not a perfect entangler when one of the
two inequalities \eqref{alphaineq1} and \eqref{alphaineq2} is not
satisfied.  We shall treat the violation of each of these
inequalities separately.

\begin{figure}
\begin{center}
\epsfxsize8cm \centerline{\epsfbox{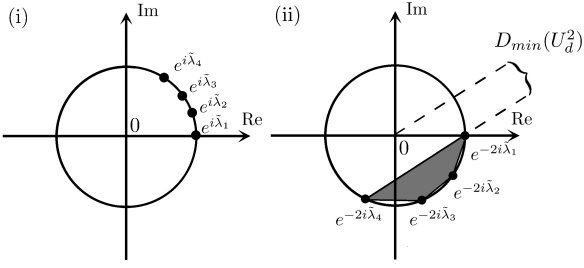}}
\end{center}
\caption{Depiction  of (i) the angles ${\tilde\lambda}_{j}$ and (ii)
the rotated eigenvalues of $U_{d}^{2}$ when $U_{AB}$ is not a
perfect entangler and inequality \eqref{alphaineq1} is violated. In
(i), all four angles lie in the first quadrant, which implies that
the product entangling capacity is given by Eq. \eqref{sini}. In
(ii), the four eigenvalues $e^{-2i{\tilde\lambda}_{j}}$ lie in the
lower half-plane, with the greatest separation being between
$e^{-2i\tilde{\lambda}_{4}}$ and $e^{-2i\tilde{\lambda}_{1}}$. The
minimum overlap is the distance from the origin to the midpoint of
the chord joining $e^{-2i{\tilde\lambda}_{1}}$ and
$e^{-2i{\tilde\lambda}_{4}}$, which gives Eq. \eqref{cosi} and thus
Eq. \eqref{maintheorem}.} \label{figure2}
\end{figure}

To prove Eq. \eqref{maintheorem} when \eqref{alphaineq1} is
violated, let us first show that
\begin{equation}
\label{sini} C_{max}^{prod}(U_{AB})={\sin}(\tilde{\lambda}_{4}).
\end{equation}
To prove this, we see that when inequality \eqref{alphaineq1} is
not satisfied, we have
\begin{equation}
\tilde{\lambda}_{4}=2({\alpha}_{x}+{\alpha}_{y})<\frac{\pi}{2}.
\end{equation}
We see from this inequality and the angular ordering in
\eqref{tildelambdaorder} that the angles $\tilde{\lambda}_{j}$ all
lie in the half-open interval $[0,{\pi}/2)$.  All differences
between neighbouring angles must clearly lie in this interval
also.  The arrangement of these angles is depicted in Figure
\ref{figure2}(i).

The ${\sin}$ function is monotonically increasing in the interval
$[0,{\pi}/2]$.  It follows that the two angles with the greatest
separation attain the maximum in Eq. \eqref{kceqn2}. Due to the
angular ordering in \eqref{tildelambdaorder}, these are
$\tilde{\lambda}_{4}$ and $\tilde{\lambda}_{1}$.  However,
$\tilde{\lambda}_{1}=0$, so we obtain Eq. \eqref{sini}.

Let us now calculate $D_{min}(U_{d}^{2})$.  Since all four angles
$2\tilde{\lambda}_{j}$ lie in first and second quadrants, the
complex numbers $e^{-2i\tilde{\lambda}_{j}}$ all lie in the lower
half-plane.  This can be seen in Figure \ref{figure2}(ii). These
complex numbers are the rotated eigenvalues of $U_{d}^{2}$, and so
$D_{min}(U_{d}^{2})$ is the distance from the origin to the convex
hull of the $e^{-2i\tilde{\lambda}_{j}}$.

From the fact that these complex numbers all lie in the lower
half-plane and from the angle ordering in \eqref{lambdaorder}, it
follows that this minimum distance is equal to the distance from the
origin to the midpoint of the chord joining
$e^{-2i\tilde{\lambda}_{4}}$ and $e^{-2i\tilde{\lambda}_{1}}$. By
elementary trigonometry, we find that
\begin{equation}
\label{cosi} D_{min}(U_{d}^{2})={\cos}(\tilde{\lambda}_{4}).
\end{equation}
Making use of both this and Eq. \eqref{sini}, we see that Eq.
\eqref{maintheorem} is satisfied.  This completes the proof of
\eqref{maintheorem} when inequality \eqref{alphaineq1} is not
satisfied.

\begin{figure}
\begin{center}
\epsfxsize8cm \centerline{\epsfbox{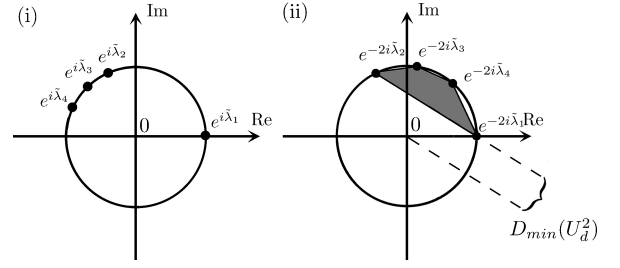}}
\end{center}
\caption{Depiction  of (i) the angles ${\tilde\lambda}_{j}$ and
(ii) the rotated eigenvalues of $U_{d}^{2}$ when $U_{AB}$ is not a
perfect entangler and inequality \eqref{alphaineq2} is violated.
In (i), $e^{i{\tilde\lambda}_{1}}=1$ and the other
$e^{i{\tilde\lambda}_{j}}$ lie in the second quadrant.  From this,
we find that the product entangling capacity is given by Eq.
\eqref{sinii}. In (ii), the four eigenvalues
$e^{-2i{\tilde\lambda}_{j}}$ lie in the upper half-plane, with the
greatest separation being between $e^{-2i\tilde{\lambda}_{1}}$ and
$e^{-2i\tilde{\lambda}_{2}}$. The minimum overlap is the distance
from the origin to the midpoint of the chord joining
$e^{-2i{\tilde\lambda}_{1}}$ and $e^{-2i{\tilde\lambda}_{2}}$,
which gives Eq. \eqref{cosii} and thus  Eq. \eqref{maintheorem}.}
\label{figure3}
\end{figure}

Let us now turn our attention to the situation where inequality
\eqref{alphaineq2} is violated.   We will begin by showing that
when this is the case, we have
\begin{equation}
\label{sinii} C_{max}^{prod}(U_{AB})={\sin}(\tilde{\lambda}_{2}).
\end{equation}
To see this, we note that when \eqref{alphaineq2} is not
satisfied, we have
\begin{equation}
\tilde{\lambda}_{2}=2({\alpha}_{y}+{\alpha}_{z})>\frac{\pi}{2}.
\end{equation}
Since $\tilde{\lambda}_{j}{\in}[0,{\pi}]$, we see that
$\tilde{\lambda}_{1}=0$ and the remaining angles lie in the second
quadrant.  This is shown in Figure \ref{figure3}(i). Let us now
consider differences between these angle differences
$\tilde{\lambda}_{j}-\tilde{\lambda}_{j'}$.  Because, from Eq.
\eqref{kceqn2}, we wish to find the maximum absolute value of the
${\sin}$, we may, without loss of generality, take $j>j'$. In the
case where $j'=1$, Eq. \eqref{sinii} follows readily by combining
the fact that the ${\sin}$ function is decreasing in the interval
$[{\pi}/2,{\pi}]$ with the angular ordering in
\eqref{tildelambdaorder}.
For $j'=2$ or $3$, let us write
$\tilde{\lambda}_{j}-\tilde{\lambda}_{j'}={\Delta}$.  We shall now
use the elementary ${\sin}$ difference rule in the following way:
\begin{equation}
\label{sindiff}
{\sin}(\tilde{\lambda}_{2})-{\sin}({\Delta})=2{\cos}\left(\frac{\tilde{\lambda}_{2}+{\Delta}}{2}\right){\sin}\left(\frac{\tilde{\lambda}_{2}-{\Delta}}{2}\right){\geq}0.
\end{equation}
The reason why we have the inequality here is as follows. Firstly,
the ${\cos}$ factor must be non-negative.  This is because
$\tilde{\lambda}_{2}+{\Delta}$ is equal to either
$\tilde{\lambda}_{3}$ or $\tilde{\lambda}_{4}$ which we know, from
Eq. \eqref{tildelambdaorder}, lies in the interval $[0,{\pi}]$.
Therefore, the argument of the ${\cos}$ factor lies in the first
quadrant and the ${\cos}$ factor is therefore non-negative. As for
the ${\sin}$ factor, we note that the angle
$(\tilde{\lambda}_{2}-{\Delta})/2$ also lies in the first quadrant,
as a consequence of Eqs. \eqref{alphaineq2} and
\eqref{tildelambdaorder}, where the ${\sin}$ function is also
non-negative.  Finally, because both terms on the left-hand side of
Eq. \eqref{sindiff} are non-negative, and because
$\tilde{\lambda}_{2}=\tilde{\lambda}_{2}-\tilde{\lambda}_{1}$ (by
virtue of the fact that $\tilde{\lambda}_{1}=0$), we obtain Eq.
\eqref{sinii}.

Let us now calculate $D_{min}(U_{d}^{2})$. From the above
considerations it follows that the $e^{-2i\tilde{\lambda}_{j}}$, the
rotated eigenvalues of $U_{d}^{2}$, all lie in the upper half-plane,
as shown in Figure \ref{figure3}(ii).   The minimum distance from 0
to the convex hull of the $e^{-2i\tilde{\lambda}_{j}}$ is clearly
equal to the distance from the origin to the midpoint of the chord
joining $e^{-2i\tilde{\lambda}_{2}}$ and
$e^{-2i\tilde{\lambda}_{1}}$. By elementary trigonometry we find
that
\begin{equation}
\label{cosii} D_{min}(U_{d}^{2})=-{\cos}(\tilde{\lambda}_{2}).
\end{equation}
Combining this with Eq. \eqref{sinii} shows that  Eq.
\eqref{maintheorem} is satisfied.  This, together with the result in
the Appendix, completes the
proof of Eq. \eqref{maintheorem} for all two-qubit unitary operators. ${\Box}$\\

Let us now make some observations about this theorem.  The first
point to note is that in Kraus and Cirac's original presentation,
the product entangling capacities of perfect and imperfect
entanglers were treated separately.  The theorem we have proven
above unifies these two scenarios, expressing as is does the
product entangling capacity of $U_{AB}$ with the single equation
Eq. \eqref{maintheorem} rather than the two equations
\eqref{kceqn0} and \eqref{kceqn1}. The two scenarios described by
Kraus and Cirac, in which $U_{AB}$ is a perfect or imperfect
entangler, correspond directly to $U_{d}$ and $U_{d}^{\dagger}$
being perfectly or imperfectly distinguishable.

The relationship in Eq. \eqref{maintheorem} between the product
entangling capacity $C_{max}^{prod}(U_{AB})$ and the
distinguishability of $U_{d}$ and $U_{d}^{\dagger}$ suggests that
$C_{max}^{prod}(U_{AB})$ may serve to quantify the non-Hermiticity
of $U_{d}$. This idea is reinforced by the following observation:
$D_{min}(U_{d}^{2})$ reaches its maximum of 1, which corresponds to
complete indistinguishability, iff $U_{d}$ is Hermitian. This is
easy to see. The distance from the origin to the midpoint of the
chord joining two points on the unit circle attains the value 1 iff
these two points are identical. Combining this with Eq.
\eqref{maintheorem} and the fact that, for imperfect entanglers, the
points in question are the (rotated) eigenvalues of $U_{d}^{2}$ that
are furthest apart, we see that this scenario arises only when all
the ${\tilde\lambda}_{j}$ are all equal, indeed equal to 0 since
${\tilde\lambda}_{1}=0$. This is equivalent to $U_{d}^{2}=1$ which
implies that $U_{d}=U_{d}^{\dagger}$ as a consequence of unitarity.
So, when the product entangling capacity of $U_{AB}$ is zero,
$U_{d}$ is Hermitian.  Also, the converse of this statement follows
trivially from Eq. \eqref{maintheorem}. So, we are led to see that
the product entangling capacity of $U_{AB}$ is directly related to
and increases with the non-Hermiticity of $U_{d}$, as quantified
operationally by our practical ability to distinguish this operator
from its adjoint with a probe state.

Finally, we note that combining Eq. \eqref{maintheorem} with Eqs.
\eqref{kceqn0} and  \eqref{kceqn1} gives the following
relationship between $C_{max}(U_{AB})$, the entangling capacity
without the restriction to initial product states, and the
distinguishability of $U_{d}$ and $U_{d}^{\dagger}$:
\begin{equation}
\label{maintheorem2}
[C_{max}(U_{AB})]^{4}+[D_{min}(U_{d}^{2})]^{2}=1.
\end{equation}
We are somewhat undecided about which of the two relationships in
Eqs. \eqref{maintheorem} and \eqref{maintheorem2} should be
regarded as being more significant. What we will find, however, in
the subsequent section, is that certain expressions involving
entangling capacities, as quantified by the entropy of
entanglement, are more naturally formulated in terms of the
product state capacity rather that the corresponding capacity
without this restriction.

\section{Information theoretic interpretations of the product entropic entangling capacity}
\label{sec:4}
\renewcommand{\theequation}{4.\arabic{equation}}
\setcounter{equation}{0}
\subsection{The first-order classical capacity}
In the preceding section, we showed that the maximum concurrence
that can be obtained by acting on an initial product state of two
qubits $|{\psi}_{A}{\rangle}{\otimes}|{\psi}_{B}{\rangle}$ with a
bipartite unitary operator $U_{AB}$ is related, through Eq.
\eqref{maintheorem}, to the distinguishability of the operators
$U_{d}$ and $U_{d}^{\dagger}$.

The act of discriminating between these two operators is a
``one-shot" procedure.  However, in the theory of classical
information transmission, one is typically concerned with asymptotic
quantities that relate to very large strings of classical
information.  It is therefore interesting to enquire as to whether
or not there is a link between such quantities and the maximum
entropy of entanglement, which we may obtain from the maximum
concurrence using Eq. \eqref{entropcab}, whose significance derives
from its usefulness in asymptotic entanglement processing.

In this section, we shall see that relationships of this nature do
indeed exist.  We shall describe two of them.  The first does not
make use of Eq. \eqref{maintheorem} and relates to the so-called
first-order classical capacity.  This capacity is derived from the
assumption that collective decoding of the signal states is
impossible.   Interestingly, in contrast with Eq.
\eqref{maintheorem}, this relationship expresses a certain
connection between the product entropic entangling capacity of
$U_{AB}$ and the {\em in}distinguishability of states, in terms of a
trade-off with the ability of certain quantum channels to faithfully
transmit classical information, as quantified by the first-order
classical capacity.   This relationship depends upon the ability to
conjugate states.  Since complex conjugation cannot be carried out
on an unknown state, the sender is required to know the state in
advance of transmission if this relationship is to correspond to a
physically realisable scenario.  However, to our knowledge, it has
not appeared in the literature so far to we include it for the sake
of completeness, as its simplicity suggests that it may be of some
formal interest.

 The second relationship we shall describe, however, does not
 suffer from this weakness.  Moreover, this second relationship,
 unlike the previous one, does make explicit use of Eq. \eqref{maintheorem} and the
connection between
 entangling capacity and distinguishability it expresses.  It
 relates to the HJSWW classical capacity \cite{HJSWW}.  Unlike the first-order capacity, the HJSWW capacity refers to situations where arbitrary collective decoding of the signal states is permitted.  Our second relationship establishes that the product entropic entangling capacity of $U_{AB}$ is precisely equal to the
 maximum of the HJSWW capacities of a certain set of quantum channels constructed using $U_{d}$ and
$U_{d}^{\dagger}$.

To describe the first of these relationships, let us recall Eq.
\eqref{entropcab}, which expresses the entropy of entanglement in
terms of the concurrence.  Since these functions are monotonically
increasing with one another, we easily conclude that the maximum
entropy of entanglement that $U_{AB}$ can generate from an initial
product state is
\begin{equation}
\label{emax1}
E^{prod}_{max}(U_{AB})=h\left(\frac{1+\sqrt{1-[C_{max}^{prod}(U_{AB})]^{2}}}{2}\right).
\end{equation}
Consider one party Alice who wishes to send classical information to
her distant colleague Bob using nonorthogonal quantum states. Alice
sends him a stream of qubits over a noiseless quantum channel. Each
of these systems acts as a signal carrier and they are all prepared
in one of the $N$ quantum states ${\rho}_{j}$, where
$j=1,{\ldots},N$. For a given signal carrier, the probability of her
preparing the state ${\rho}_{j}$ is $p_{j}$. On receiving the state
${\rho}_{j}$, Bob's task is to determine, using a measurement, which
of these states Alice sent. The possible states may be nonorthogonal
so, in general, Bob will not be able to distinguish among them
reliably.

We assume here that Bob measures each of the signal carriers
individually.  That is to say that collective measurements on
strings of signal carriers are not permitted.  Subject to this
restriction, the most general measurement he can perform will have
$K$ possible outcomes, for some positive integer $K$.
Corresponding to outcome $k{\in}[1,{\ldots},K]$ is a positive
operator ${\Pi}_{k}$ acting on the Hilbert space of a single
signal carrier.   These operators, known as positive,
operator-valued measure (POVM) elements, must sum to the identity
operator on the single signal carrier Hilbert space.

These operators serve to describe the statistical properties of
Bob's measurement.  The communication channel is characterised by
the channel matrix.  The elements of this matrix are the
probabilities of obtaining each of the $K$ possible measurement
results for each of the $N$ possible initial states.  These
probabilities are
\begin{equation}
\label{POVMprob} P(k|{\rho}_{j})=\mathrm{Tr}({\Pi}_{k}{\rho}_{j}).
\end{equation}
For nonorthogonal signal states, the channel matrix cannot be the
identity matrix.  It will have nonzero off-diagonal elements and so
the channel behaves in many respects like a noisy classical channel,
where the noise is an unavoidable consequence of the
indistinguishability of nonorthogonal quantum states.  The
similarity is sufficiently strong to enable us to employ Shannon's
noisy coding theorem.  By redundant coding of the classical messages
at Alice's end and classical error correction of the measurement
results at Bob's, Alice can send Bob asymptotically error-free
classical information at a nonzero rate.  The maximum rate at which
she can do this is characterised by the so-called first-order
classical capacity. This quantity is defined in terms of the mutual
information
\begin{equation}
I=\sum_{k=1}^K \sum_{j=1}^N
        p_{j}P(k|{\rho}_{j})\log \left( \frac{P(k|{\rho}_{j})}
        { \sum_{j'=1}^N p_{j'} P(k|{\rho}_{j'})} \right).
\end{equation}
The mutual information clearly depends not only on the states
${\rho}_{j}$ and their probabilities $p_{j}$, but also, through
the POVM elements in Eq. \eqref{POVMprob}, on Bob's measurement.

The significance of this quantity is as follows: the maximum rate at
which Alice can send Bob asymptotically error-free classical
information using the quantum states ${\rho}_{j}$, over all possible
probability distributions $p_{j}$ and all noncollective measurements
that Bob may perform, is given by ${\cal C}_{1}(\{{\rho}_{j}\})$
bits per signal carrier.  This quantity is the first-order classical
capacity, and it is given by
\begin{equation}
{\cal C}_{1}(\{{\rho}_{j}\})=\max_{\{p_{j}\}}\max_{\{\Pi_{k}\}}I,
\end{equation}
that is, it is the maximum of the mutual information with respect
to the probabilities $p_{j}$ and the measurement Bob performs.  We
should mention that implicit in the maximisation over the set of
possible measurements is a maximisation over $K$, the number of
measurement outcomes.

This quantity is difficult to calculate for most sets of quantum
states. However, for a pair of pure states $|{\psi}_{1}{\rangle}$
and $|{\psi}_{2}{\rangle}$, it can be calculated exactly. Sasaki et
al. \cite{Sasaki1,Sasaki2} have shown that the explicit expression
is

\begin{equation}
\label{cone} {\cal
C}_{1}(\{|{\psi}_{j}{\rangle}\})=1-h\left(\frac{1+\sqrt{1-|{\langle}{\psi}_{1}|{\psi}_{2}{\rangle}|^{2}}}{2}\right).
\end{equation}
Notice the similarity between the second term in this
 the expression and that we gave for the product entropic entangling capacity in Eq. \eqref{emax1}. Clearly, this similarity
would be even greater if we could interpret the concurrence as an
overlap between two pure states.  Fortunately, we are able to do
this.  Using the results of Wootters \cite{Wootters}, it was
observed by Leifer et al. \cite{LHL} that the concurrence of a
bipartite pure state $|{\Psi}{\rangle}$ of two qubits can be written
as
\begin{equation}
C(|{\Psi}{\rangle})=|{\langle}{\Psi}|{\sigma}_{y}{\otimes}{\sigma}_{y}|{\Psi}^{*}{\rangle}|.
\end{equation}
Here, * denotes complex conjugation in the computational basis,
i.e. the basis of product eigenstates of
${\sigma}_{z}{\otimes}{\sigma}_{z}$.  We note that, in this basis, $U_{d}^{*}=U_{d}^{\dagger}$.

\begin{figure}
\begin{center}
\epsfxsize8cm \centerline{\epsfbox{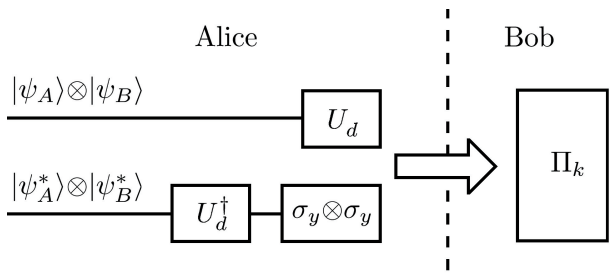}}
\end{center}
\caption{Illustration of a scenario that leads to an
interpretation of the product entropic entangling capacity of
$U_{AB}$ in terms of the first-order classical capacity.  A
classical description of the product state
$|{\psi}_{A}{\rangle}{\otimes}|{\psi}_{B}{\rangle}$ is used to
manufacture, at will, either this state or
$|{\psi}_{A}^{*}{\rangle}{\otimes}|{\psi}_{B}^{*}{\rangle}$, its
complex conjugate in the computational basis.  These states are
then acted on by $U_{d}$ and
$({\sigma}_{y}{\otimes}{\sigma}_{y})U_{d}^{\dagger}$ respectively.
The resulting signals and then received by Bob who measures them
individually.  The maximum rate at which Alice can send classical
information to Bob this way, over all product states
$|{\psi}_{A}{\rangle}{\otimes}|{\psi}_{B}{\rangle}$, is related to
the product entropic entangling capacity of $U_{AB}$ through Eq.
\eqref{relation1}.} \label{figure4}
\end{figure}

So let us make the identifications
\begin{eqnarray}
\label{psi1def}
|{\psi}_{1}{\rangle}&=&|{\Psi}{\rangle}=U_{d}|{\psi}_{A}{\rangle}{\otimes}|{\psi}_{B}{\rangle},\\
\label{psi2def}
|{\psi}_{2}{\rangle}&=&({\sigma}_{y}{\otimes}{\sigma}_{y})U_{d}^{\dagger}|{\psi}_{A}^{*}{\rangle}{\otimes}|{\psi}_{B}^{*}{\rangle},
\end{eqnarray}
for some product state
$|{\psi}_{A}{\rangle}{\otimes}|{\psi}_{B}{\rangle}{\in}{\cal
H}^{{\otimes}2}$. Using these expressions, we now describe an
information transmission procedure that links the product entropic
entangling capacity of $U_{AB}$ with the first-order classical
capacity of certain quantum channels.  The situation we shall
consider is depicted in Figure \ref{figure4}.

Each of Alice's transmissions begins with a classical description
of some product state
$|{\psi}_{A}{\rangle}{\otimes}|{\psi}_{B}{\rangle}$. To prepare
the state $|{\psi}_{1}{\rangle}$, she creates
$|{\psi}_{A}{\rangle}{\otimes}|{\psi}_{B}{\rangle}$ and then
applies the unitary operator $U_{d}$. To prepare the state
$|{\psi}_{2}{\rangle}$, she creates the state
$|{\psi}_{A}^{*}{\rangle}{\otimes}|{\psi}_{B}^{*}{\rangle}$ and
then applies the unitary operator
$({\sigma}_{y}{\otimes}{\sigma}_{y})U^{\dagger}_{d}$. Notice the
necessity of her having a classical description of the state
$|{\psi}_{A}{\rangle}{\otimes}|{\psi}_{B}{\rangle}$ here.  To
produce the state $|{\psi}_{2}{\rangle}$, she cannot begin with
the an unknown product state
$|{\psi}_{A}{\rangle}{\otimes}|{\psi}_{B}{\rangle}$ and then apply
a quantum operation to it, as no quantum operation can conjugate
an unknown state.

Following Alice's preparation of her desired state
$|{\psi}_{1}{\rangle}$ or $|{\psi}_{2}{\rangle}$, she sends it to
Bob, who proceeds to extract classical information.  We find, upon
combining Eqs. \eqref{emax1} and \eqref{cone} with the fact that
product entropic entangling capacities of $U_{AB}$ and $U_{d}$ are
equal, that the product entropic entangling capacity of $U_{AB}$
and the first-order classical capacity of the channel are related
by

\begin{equation}
\label{relation1} E_{max}^{prod}(U_{AB}
)+\max_{|{\psi}_{A}{\rangle}{\otimes}|{\psi}_{B}{\rangle}{\in}{\cal
H}^{{\otimes}2}}{\cal C}_{1}(\{|{\psi}_{j}{\rangle}\})=1,
\end{equation}
where the $|{\psi}_{j}{\rangle}$ are defined by Eqs.
\eqref{psi1def} and \eqref{psi2def}. This relationship clearly
expresses a trade-off between the product entropic entangling
capacity of $U_{AB}$ and the maximum first-order classical
capacity over the channels we have have described.

\subsection{The HJSWW classical capacity}

The relationship we have just described depends upon the
conjugation of states. As a such, a classical description of the
product state $|{\psi}_{A}{\rangle}{\otimes}|{\psi}_{B}{\rangle}$
is required. Despite this, the simple nature of this relationship
is something that we believe is of some formal interest.

Here we shall describe a relationship between the product entropic
entangling capacity and classical capacities of quantum channels
that does not suffer from this weakness. Also, this relationship
does make use of Eq. \eqref{maintheorem}, unlike the previous one.
Finally, this relationship involves the HJSWW classical capacity of
quantum channels.  This capacity, which allows for collective
measurements, is at least as high as the first-order classical
capacity.

Let us once again consider Eq. \eqref{maintheorem}.  We see from
this equation and Eq. \eqref{emax1} that the product entropic
entangling capacity can be expressed as
\begin{equation}
\label{emax2}
E^{prod}_{max}(U_{AB})=h\left(\frac{1+D_{min}(U^{2}_{d})}{2}\right).
\end{equation}

The relationship we shall describe makes use of the fact that if
Alice wishes to send classical information to Bob, but unlike in
the previous scenario, he is able to perform collective
measurements on the signal carriers he receives, then an enhanced
rate of classical information transmission can be achieved.  This
effect, known as the superadditive quantum coding gain, has
recently been demonstrated for the first time in the laboratory by
Sasaki and collaborators \cite{Fujiwara,Takeoka}.

Suppose Alice wishes to send classical information to Bob by
sending, over a noiseless quantum channel, states drawn from a
source of $N$ pure states $|{\psi}_{j}{\rangle}$ with respective
probabilities $p_{j}$.  In the limit when Bob can perform
measurements on arbitrarily long strings of signal carriers, the
classical capacity is given by \cite{HJSWW}
\begin{equation}
{\cal
C}_{\infty}(\{|{\psi}_{j}{\rangle}\})=\max_{\{p_{j}\}}S({\rho}),
\end{equation}
where $S$ is the von Neumann entropy and
${\rho}=\sum_{j=1}^{N}p_{j}|{\psi}_{j}{\rangle}{\langle}{\psi}_{j}|$.
For two pure states, $|{\psi}_{1}{\rangle}$ and
$|{\psi}_{2}{\rangle}$, the explicit form of the capacity ${\cal
C}_{\infty}(\{|{\psi}_{j}{\rangle}\})$ is easily determined. We
find
\begin{equation}
S({\rho})=h\left(\frac{1+\sqrt{(p_{1}-p_{2})^{2}+4p_{1}p_{2}|{\langle}{\psi}_{1}|{\psi}_{2}{\rangle}|^{2}}}{2}\right).
\end{equation}
Maximising this expression with respect to the probabilities
$p_{j}$ and taking into account the normalisation constraint
$p_{1}+p_{2}=1$ leads to the conclusion that the maximum is
obtained when both probabilities are equal to $1/2$.  This gives
\begin{equation}
\label{hswcapacity}
 {\cal
C}_{\infty}(\{|{\psi}_{j}{\rangle}\})=h\left(\frac{1+|{\langle}{\psi}_{1}|{\psi}_{2}{\rangle}|}{2}\right).
\end{equation}
This expression is highly reminiscent of Eq. \eqref{emax2},
especially in view of the fact that $D_{min}(U_{d}^{2})$ is an
overlap.

\begin{figure}
\begin{center}
\epsfxsize8cm \centerline{\epsfbox{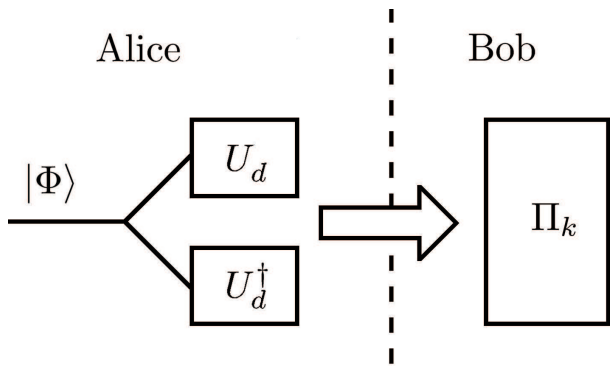}}
\end{center}
\caption{Illustration of a scenario that leads to an
interpretation of the product entropic entangling capacity of
$U_{AB}$ in terms of the HJSWW classical capacity.  Alice produces
a number of copies of a bipartite pure state $|{\Phi}{\rangle}$.
Upon each one, she acts with either $U_{d}$ or $U_{d}^{\dagger}$.
The resulting modified states are sent to Bob, who is able to
perform collective measurements on several of them together.  The
maximum rate at which Alice can send classical information to Bob
this way, over all possible initial states $|{\Phi}{\rangle}$, is
equal to the product entropic entangling capacity of $U_{AB}$, as
expressed by Eq. \eqref{relation2}.} \label{figure5}
\end{figure}

To see how the similarity of these two expressions can lead to a
concrete physical relationship between the product entropic
entangling and HJSWW classical capacities,  consider the  scenario
depicted in Figure \ref{figure5}. In Alice's laboratory there is a
machine that produces many copies of some two-qubit pure state
$|{\Phi}{\rangle}$. This machine can be set to produce any such
state.  However, it cannot be reset; once one particular state has
been set to be mass produced, it is impossible to reprogram the
machine to produce a different state.

On each copy of the state $|{\Phi}{\rangle}$, Alice acts with one
of the operators $U_{d}$ and $U_{d}^{\dagger}$. For each copy, she
is completely free in her choice of which operator to use.  She
then transmits the modified state to Bob, who is able to perform
arbitrary collective measurements on arbitrarily long strings of
signal states.

The HJSWW capacity of this channel is given by Eq.
\eqref{hswcapacity}, where
\begin{eqnarray}
\label{psi3def}
|{\psi}_{1}{\rangle}&=&U_{d}^{\dagger}|{\Phi}{\rangle}, \\
\label{psi4def}
|{\psi}_{2}{\rangle}&=&U_{d}|{\Phi}{\rangle}.
\end{eqnarray}
Clearly, the only variable is the initial state $|{\Phi}{\rangle}$
and it is interesting to maximise the HJSWW capacity with respect
to this state. It is easy to see that the maximum is, essentially
by definition, given by Eq. \eqref{emax2}. We therefore obtain
\begin{equation}
\label{relation2} E^{prod}_{max}(U_{AB}
)=\max_{|{\Phi}{\rangle}{\in}{\cal H}^{{\otimes}2}}{\cal
C}_{\infty}(\{|{\psi}_{j}{\rangle}\}),
\end{equation}
where the $|{\psi}_{j}{\rangle}$ are defined by Eqs.
\eqref{psi3def} and \eqref{psi4def}.

We regard this relationship as being the asymptotic analogue of
Eq. \eqref{maintheorem}.  In Eq. \eqref{maintheorem}, we saw that
the more distinguishable $U_{d}$ and $U_{d}^{\dagger}$ are, the
higher is the product entangling capacity of $U_{AB}$ in terms of
the concurrence.  It is natural then that, in the asymptotic
limit, the entangling capacity, which is quantified by the maximum
entropy of entanglement, is related to the extent to which we can
send classical information by modulating an initial state with
$U_{d}$ and $U_{d}^{\dagger}$, which is precisely what is
expressed by Eq. \eqref{relation2}.

\section{Discussion}
\label{sec:5}
\renewcommand{\theequation}{4.\arabic{equation}}
\setcounter{equation}{0}

The principal aim of the present paper has been to describe a
interesting relationship between the entangling capacity of a
two-qubit unitary operator $U_{AB}$ and the distinguishability of
its canonical form $U_{d}$ and its adjoint $U_{d}^{\dagger}$.  We
saw, in Eq. \eqref{maintheorem}, that the product entangling
capacity of $U_{AB}$ quantifies the distinguishability of $U_{d}$
and $U_{d}^{\dagger}$. This implies that it may also be viewed as a
measure of the non-Hermiticity of $U_{d}$.  As a measure of
non-Hermiticity, it is highly significant from an practical point of
view, since it quantifies the extent to which we can operationally
distinguish $U_{d}$ from $U_{d}^{\dagger}$ with a probe state.

In the asymptotic limit, the entanglement properties of pure
bipartite states are naturally quantified using the entropy of
entanglement.  It is interesting to enquire as to whether or not our
relationship between the maximum concurrence and minimum overlap has
any implications for this limit.  Indeed it does.  We found two
relationships between the product entropic entangling capacity and
the classical capacities of quantum channels.  The first does not
actually involve Eq. \eqref{maintheorem} but we included it for the
sake of completeness as, to our knowledge, this relationship was not
previously known.  It also suffers from a weakness; it corresponds
to a classical information transmission scenario where the sender,
Alice, must must know the state in advance.

Our second relationship, however, is more satisfactory in a number
of respects.  Firstly, it makes explicit use of Eq.
\eqref{maintheorem}.  Secondly, it applies to an unknown initial
state.  Finally, it corresponds to a more general scenario where
collective decoding of the signal carriers is permitted, thus making
greater use of the ability of quantum states to carry classical
information.

The results in this paper relate to bipartite unitary operators
where the subsystems are qubits.  It is natural to enquire as to
whether or not they can be extended to higher dimensional
subsystems. What is interesting is the fact that the theory of
distinguishing between a pair of unitary operators outlined in
Section \ref{sec:2} applies very generally, certainly to all unitary
operators on finite dimensional Hilbert spaces. However, the theory
of the entangling capacity of bipartite unitary operators is not
similarly well-developed.  Exact results are only known for qubit
subsystems.  It should be said in this context that the entangling
power, which is the {\em average} amount of entanglement that an
arbitrary bipartite unitary operator can generate, has been
established by Zanardi and co-workers.
\cite{Zanardi1,Zanardi2,Zanardi3} for higher-dimensional subsystems.
However, here we are interested in the entangling capacity, which is
the {\em maximum} amount of entanglement that can be  created.

One of the reasons why the entangling capacity of a general
$D_{1}{\times}D_{2}$ unitary operator, for $D_{1},D_{2}>2$, has
not yet been established is the fact that a suitable canonical
form $U_{d}$ is not known for such operators.  Also, for higher
dimensional subsystems, there are difficulties in defining a
single quantity which generalises the concurrence
\cite{BDHHH,Gour}. This is related to fact that for higher
dimensional bipartite pure entangled states, there is no single
quantity which completely characterises the entanglement.  In
general, we have a set of entanglement monotones which can vary
with a certain degree of independence from one state to another
\cite{Vidal}. For one state to be unambiguously more entangled
than another, the values of all entanglement monotones must be at
least as high for one of the states as they are for the other.
Otherwise, the states are incomparable.

These considerations lead us to conclude this paper with a
question.  Might it be the case that for $D_{1}{\times}D_{2}$
unitary operators with $D_{1},D_{2}>2$, there is, in general, no
unique entangling capacity at the one-shot level, even if we
restrict ourselves to initial product states?  This would be the
case if, for all $D_{1},D_{2}>2$, we can find a unitary operator
$U_{AB}$ which has the following property:  there is no initial
product state $|{\psi}_{A}{\rangle}{\otimes}|{\psi}_{B}{\rangle}$
such that
$U_{AB}|{\psi}_{A}{\rangle}{\otimes}|{\psi}_{B}{\rangle}$
simultaneously maximises all entanglement monotones.  Due to the
fact that these monotones exhibit a certain degree of mutual
independence, this seems possible.

\section*{ACKNOWLEDGMENTS}
This work was funded by the Science Foundation Ireland and by the
University of Hertfordshire.

\section*{APPENDIX: PROOF OF EQ. \eqref{maintheorem} FOR NEGATIVE ${\alpha}_{z}$}
\renewcommand{\theequation}{A.\arabic{equation}}
\setcounter{equation}{0}

We prove here that Eq. \eqref{maintheorem} holds when ${\alpha}_{z}$
is negative. This can be done rather easily if it also holds for
non-negative ${\alpha}_{z}$, and we established that it does in
Section \ref{sec:3}.  To proceed, we note that the validity of Eq.
\eqref{maintheorem} for non-negative ${\alpha}_{z}$ implies that
\begin{equation}
\label{dprimetheorem}
[C_{max}^{prod}(U_{d'})]^{2}+[D_{min}(U_{d'}^{2})]^{2}=1.
\end{equation}
for all $d'=({\alpha}_{x}',{\alpha}_{y}',{\alpha}_{z}')$ which
satisfy
\begin{equation}
\label{alphaorder2}
0{\leq}{\alpha}_{z}'{\leq}{\alpha}_{y}'{\leq}{\alpha}_{x}'{\leq}{\pi}/4.
\end{equation}
Now consider $U_{AB}$  and $U_{d}$ related by Eq. \eqref{con} where
$d=({\alpha}_{x},{\alpha}_{y},{\alpha}_{z})$, ${\alpha}_{z}<0$ and
inequality \eqref{alphaorder0} holds.  We make the following
observation: if ${\alpha}_{x}'={\alpha}_{x}$,
${\alpha}_{y}'={\alpha}_{y}$ and ${\alpha}_{z}'=-{\alpha}_{z}$, then
\begin{equation}
\label{Udtrans}
({\sigma}_{z}{\otimes}1)U_{d}({\sigma}_{z}{\otimes}1)=U_{d'}^{\dagger}.
\end{equation}

This implies that $U_{AB}$ is locally equivalent to and therefore
has the same product entangling capacity as $U_{d'}^{\dagger}$.
Indeed, it also has the same product entangling capacity as
$U_{d'}$, because $U_{d'}^{\dagger}$ is simply the complex
conjugate of $U_{d'}$ in the computational basis.  It has been
established \cite{KC} that complex conjugation in this basis does
not change the entangling capacity.  We may therefore write
\begin{equation}
\label{Ceq}
 C_{max}^{prod}(U_{AB})=C_{max}^{prod}(U_{d'}).
\end{equation}
Also, the fact that $U_{d}$ and $U_{d'}^{\dagger}$ are unitarily
related through Eq. \eqref{Udtrans} implies that
$D_{min}(U_{d}^{2})=D_{min}(U_{d'}^{{\dagger}2})$, which is easily
seen to be also equal to $D_{min}(U_{d'}^{2})$.  We may then write
\begin{equation}
\label{Deq} D_{min}(U_{d}^{2})=D_{min}(U_{d'}^{2}).
\end{equation}
Finally, substitution of Eqs. \eqref{Ceq} and \eqref{Deq} into Eq.
\eqref{dprimetheorem} gives Eq. \eqref{maintheorem} as desired.

\end{document}